\documentclass[dvips]{acta}
\usepackage{supertabular,lscape,epsfig}
\usepackage{amssymb}
\usepackage{amsmath}
\SetPages{1}{7}
\SetVol{66}{2016}

\begin{document}

\begin{Titlepage}

\Title { On the Light Curves of AM CVn }

\Author {J.~~S m a k}
{N. Copernicus Astronomical Center, Polish Academy of Sciences,\\
Bartycka 18, 00-716 Warsaw, Poland\\
e-mail: jis@camk.edu.pl }

\Received{  }

\end{Titlepage}

\Abstract { Light curves of AM CVn are analyzed by decomposing them into their 
Fourier components. 
The amplitudes of the fundamental mode and overtones of the three components: 
the superhumps, the negative superhumps and the orbital variations, are found 
to be variable. 
This implies that variations in the shape of the observed light curve 
of AM CVn are not only due to the interference between those components, 
but also due to the variability of their parameters. 
} 
{binaries: cataclysmic variables, stars: individual: AM CVn }

\section { Introduction } 

AM CVn is a prototype of the ultra short period, helium cataclysmic binaries 
(Nelemans 2005, Kotko et al. 2012).  
Its light variations, discovered in 1962 (Smak 1967), are so complicated 
that it took many decades and several, extensive photometric   
and spectroscopic studies (Patterson et al. 1993, Harvey et al. 1998, 
Skillman et al. 1999, Nelemans et al. 2001, Roelofs et al. 2006 and references therein) 
before they were fully interpreted and the basic binary system parameters 
well established. 

Thanks to those investigations the light variations of AM CVn are now known 
to be a superposition of three components: 
the superhumps with $P_{SH}=1051.2s$, the negative superhumps with 
$P_{nSH}=1011.4s$, and variations with the orbital period $P_{orb}=1028.7s$. 
The superhumps are the dominant component and their period is the main 
observed period. 
In view of this it could be added that light variations of AM CVn, when discovered 
in 1962, were the first -- unrecognized at that time(!) -- example of superhumps. 

There are still problems requiring further attention, such as  
the superhump period variations (cf. Fig.2 in Skillman et al. 1999), 
or problems related to the fact that the observed light curve has the shape 
of a distorted double sine-wave, dominated by the strong first overtone  
525s signal, which is the peculiar property of AM CVn. 

Another problem is related to large variations in the shape of the light curve, 
observed on shorter time scales. 
The aim of the present paper is to clarify this point by 
decomposing representative light curves of AM CVn into their Fourier components. 

\section { The Seasonal Mean Light Curves } 

Skillman et al. (1999) collected and analyzed long series of photometric 
observations of AM CVn which allowed them to determine the seasonal mean 
superhump (SH), orbital (orb) and negative superhump (nSH) light curves 
observed in the years: 1978, 1997 and 1998 (Skillman et al. 1999, Figs 5 and 6). 

Those light are decomposed into their fundamental mode and the first three 
overtones 

\beq
m~=~<m>~+~\sum_{k=0}^3~A_k~\cos [2\pi~(k+1)~(\phi-\phi_{k,min})] ,
\eeq 

\noindent 
where $k=0$ corresponds to the fundamental mode, while $k=1,2,3$ -- to the overtones. 
\footnote 
{It must be noted here that Eq.(1) in Smak (2016) was -- regretfully -- 
misprinted. Its correct form, used in the analysis, was identical with 
Eq.(1)  given above. }
The resulting values of the amplitudes and phases of minimum are listed in Table 1; 
their formal errors are quite small: $\sigma_A\sim \pm 0.3-0.6$mmag 
and $\sigma_{\phi}\sim \pm 0.01-0.03$.

An important comment must be made here in order to avoid confusion and possible 
misunderstandings. In the case of the overtones the phases of minima can be defined 
in two different ways: either by refering them to the main period $P$, 
or to the {\it overtone} period: $P_k=P/(k+1)$. 
The resulting {\it fundamental mode} phases $\phi_{min}^{fm}$ and the {\it overtone} 
phases $\phi_{min}^{ovt}$ are related by 

\beq
\phi_{min}^{fm}~=~\phi_{min}^{ovt}/(k+1) . 
\eeq

\noindent
The phases defined by Eq.(1) and listed in Table 1 are -- obviously -- the 
{\it fundamental mode} phases.

\begin{table}[h!]
{\parskip=0truept
\baselineskip=0pt {
\medskip
\centerline{Table 1}
\medskip
\centerline{ Fourier Components of the Seasonal Mean and 1974 Light Curves }
\medskip
$$\offinterlineskip \tabskip=0pt
\vbox {\halign {\strut
\vrule width 0.5truemm #&	
\quad\hfil#\quad&               
\vrule#&			
\quad\hfil#\quad&               
\vrule#&			
\quad\hfil#\quad&               
\vrule#&			
\quad\hfil#\quad&               
\vrule#&			
\quad\hfil#\quad&               
\vrule width 0.5 truemm # \cr	
\noalign {\hrule height 0.5truemm}
&&&&&&&&&&\cr
&A(mmag)&& $A_0$ && $A_1$  &&$A_2$&&$A_3$&\cr
&&&&&&&&&&\cr
\noalign {\hrule height 0.2truemm}
&&&&&&&&&&\cr
& SH <1978> && 1.7 && 10.6 && 2.5 && 0.8 &\cr
&    <1997> && 2.3 &&  9.7 && 3.3 && 0.9 &\cr
&    <1998> && 3.0 && 11.8 && 3.0 && 1.4 &\cr
& 1974\quad && 7.5 && 11.5 && 1.6 && 3.5 &\cr   
&&&&&&&&&&\cr
&orb <1997> && 4.1 &&  1.6 && 0.4 && 0.4 &\cr
&    <1998> && 4.0 &&  2.7 && 0.2 && 0.6 &\cr
&   1974~~  && 2.8 &&  8.2 && 4.8 && 0.6 &\cr   
&&&&&&&&&&\cr
&nSH <1978> && 6.3 &&  0.6 && 0.6 && 0.3 &\cr
&    <1998> && 9.0 &&  0.4 && 0.5 && 0.4 &\cr
&   1974~   && 5.8 &&  4.4 && 1.6 && 0.6 &\cr   
&&&&&&&&&&\cr
\noalign {\hrule height 0.2truemm}
&&&&&&&&&&\cr
&Phases \hfil&&$\phi_{0,min}$&&$\phi_{1,min}$&&$\phi_{2,min}$&&$\phi_{3,min}$&\cr
&&&&&&&&&&\cr
\noalign {\hrule height 0.2truemm}
&&&&&&&&&&\cr
& SH <1978> && -0.15 && +0.03 && -0.04 && -0.04 &\cr
&    <1997> && -0.31 && +0.04 && -0.02 && +0.02 &\cr
&    <1998> && -0.25 && +0.04 && -0.04 && -0.02 &\cr
& 1974\quad && -0.04 && +0.02 && +0.03 && -0.03 &\cr   
&&&&&&&&&&\cr
&orb <1997> && -0.04 &&  0.00 && +0.10 && -0.11 &\cr
 &    <1998> && -0.03 && +0.02 && -0.06 && +0.09 &\cr
& 1974\quad && +0.01 && +0.02 && -0.05 && +0.07 &\cr   
&&&&&&&&&&\cr
&nSH <1978> && -0.01 && +0.17 && +0.06 && -0.05 &\cr
&    <1998> &&  0.00 && +0.12 && -0.12 && -0.06 &\cr
& 1974\quad && +0.02 && +0.02 && -0.11 && -0.09 &\cr   
&&&&&&&&&&\cr
\noalign {\hrule height 0.5truemm}
}}$$
}}
\end{table}

Returning to Table 1 few facts can be noted with respect to the superhump component: 
(1) The amplitudes and phases of minima of the overtones do not change 
significantly from season to season; this explains the stability 
of the light curve over longer time scales. 
(2) The amplitude of the fundamental mode is very low, but not negligible, 
while its phase of minimum is probably variable. 
(3) The phase of minimum of the dominant first overtone differs from $0$ 
and it is due to the contributions from the second and third overtones that 
in the observed curve it is shifted to $0$.

\section { The 1974 Light Curves } 

\subsection {The Light Curves } 

In order to study the evolution of the light curve of AM CVn on a short time 
scale we analyze the light curves observed on two nights in January 1974 
(Smak 1975, Figs 1 and 2). They are shown in the left panels of Figs 1 and 2.  
Each curve is based on three 1051s cycles (i.e. $3P_{SH}$) 
and each next curve is shifted with respect to the previous one by $1.5P_{SH}$. 
The phases are defined by the following local elements 

\beq
{\rm Pri.Min.}~=~JDhel~2442071.81746~+~0.01216667~\times~E~.
\eeq

\begin{figure}[htb]
\epsfysize=14.0cm 
\hspace{-1.0cm}
\epsfbox{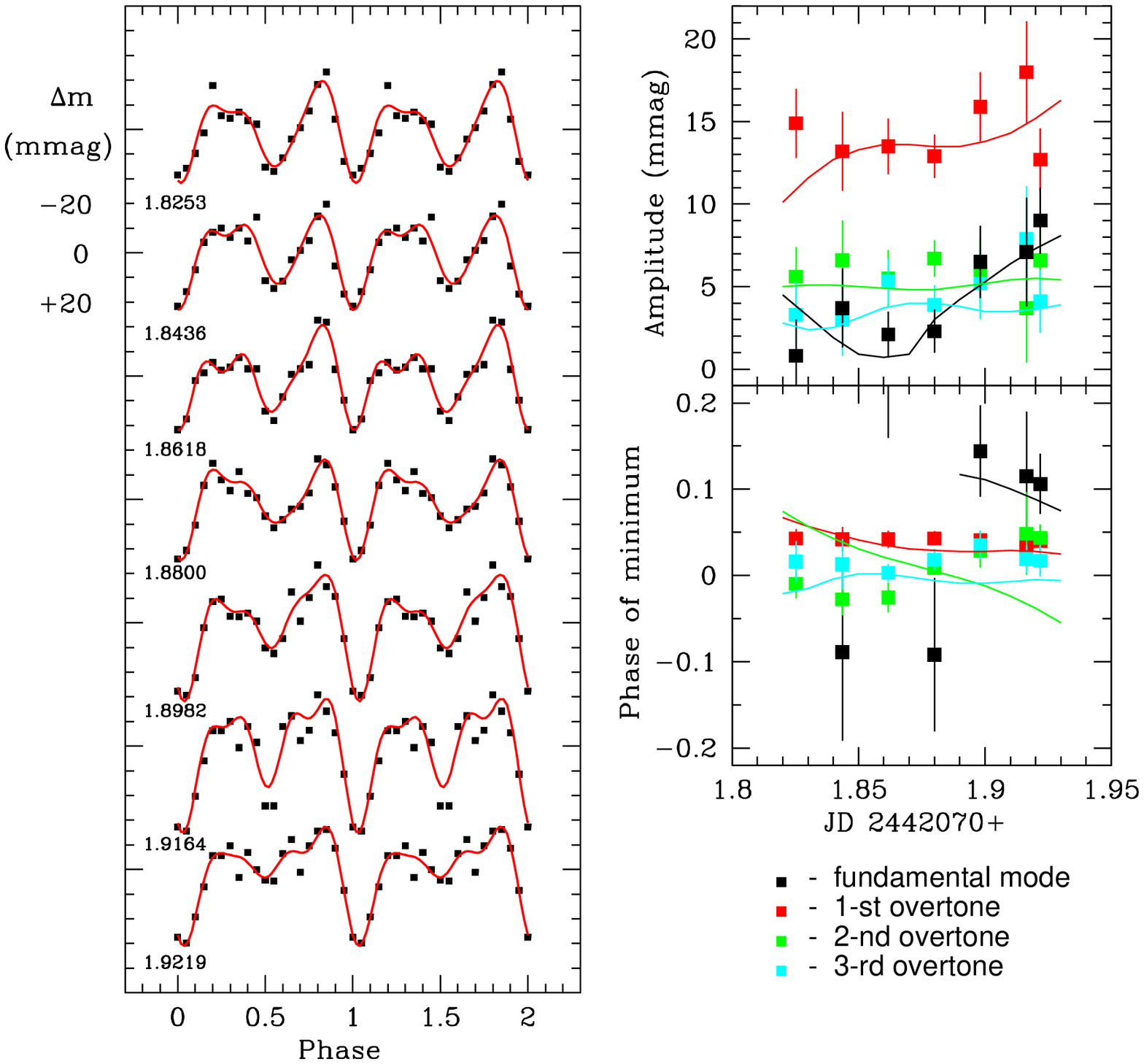} 
\vskip -10truemm
\FigCap { Analysis of light curves of AM CVn observed on January 23/24, 1974. 
{\it Left:} The light curves; each curve is based on three 1051s cycles 
and each next curve is shifted with respect to the previous one by $1.5P_{SH}$. 
Numbers below each curve give the mid-interval dates in JD-2442070. 
Red lines represent fits with four Fourier components as described in Section 3.2.  
{\it Right:} Amplitudes and phases of minimum of the fundamental mode and 
the overtones. Lines represent solution discussed in Section 3.3.}
\end{figure}

\begin{figure}[htb]
\epsfysize=14.0cm 
\hspace{-1.0cm}
\epsfbox{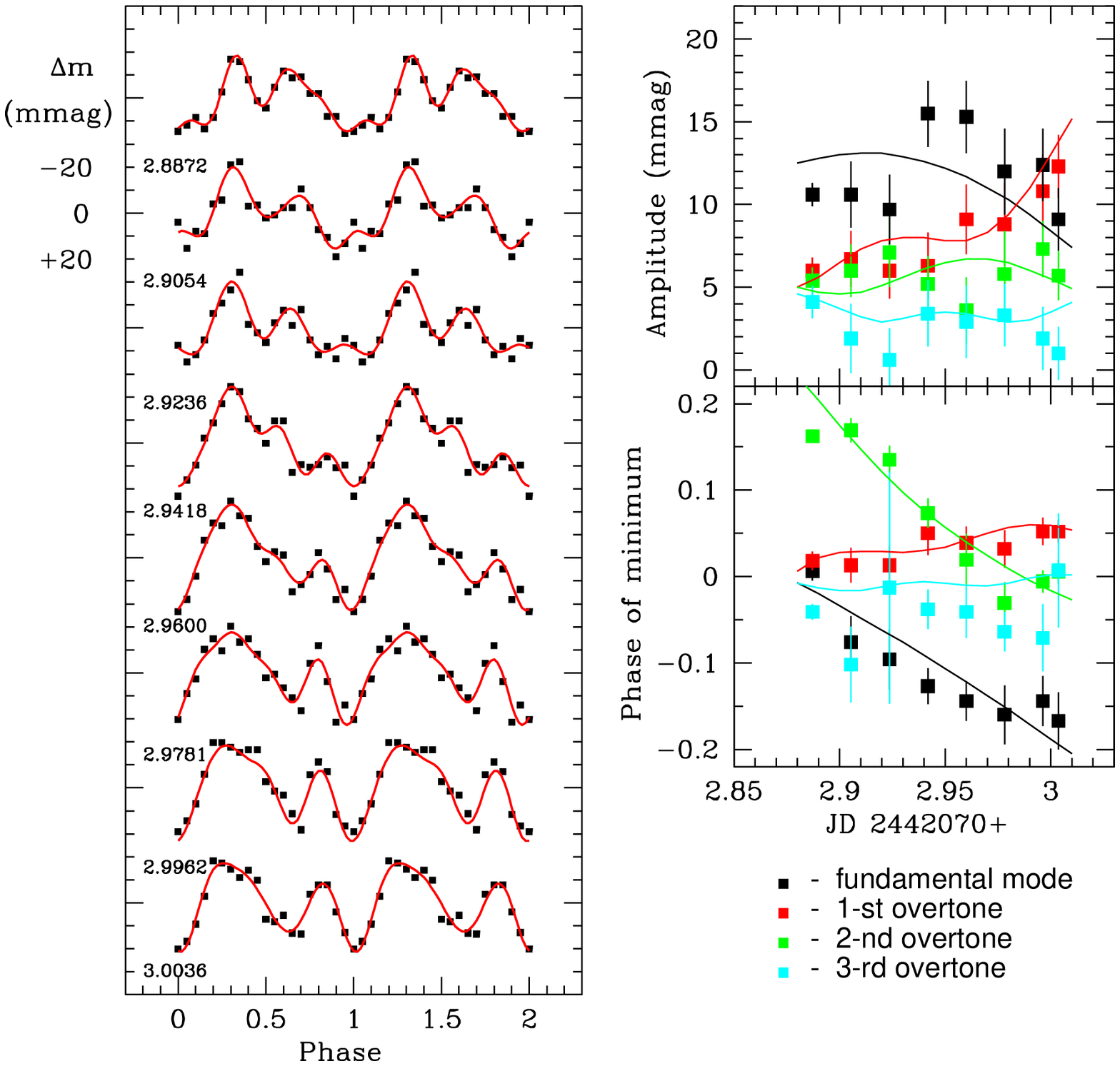} 
\vskip -10truemm
\FigCap { Analysis of light curves of AM CVn observed on January 24/25, 1974. 
See caption to Fig.1. }
\end{figure}

First of all we note that the behavior of AM CVn on those two nights was 
different. On the first night (Fig.1) the light curve was of the "standard" 
shape and only at the end of the run the depth of the secondary minimum 
started to decrease. 
On the second night (Fig.2) all light curves were peculiar with 
the mean light curve (not shown here) showing only the first minimum and the 
first maximum. 

It is obvious that the changes of the observed light curve which occur on 
a short time scale are due to the interference between its three components: 
the superhumps, the negative superhumps and the orbital variations. 
To describe their contributions one would have to determine -- for each of them -- 
the amplitudes and phases of minimum of each of the four Fourier components,  
altogether then: $3\times 4\times 2=24$ free parameters. This is obviously impossible. 
The only way to improve the situation is to begin with the Fourier analysis of the 
observed light curves.

\subsection {The Fourier Analysis }

All light curves shown in the left panels of Figs.1 and 2 were 
decomposed into their fundamental mode and the first three overtones. 
The results are shown in the right panels of those figures, where the resulting 
amplitudes and {\it fundamental mode} phases of minima are plotted as a function of time. 

All parameters show considerable variations.  
The only exception are the phases of minimum of the first overtone  
which are practically constant and equal to those obtained 
from the mean seasonal superhump light curves. This illustrates the stability 
of the main 525s signal. 

\subsection { The Solution }  

Each of the Fourier components can be represented in the form 

\beq
\Delta m_k~=~A_k^{SH}\cos \phi_k^{SH}~+~A_k^{orb}\cos \phi_k^{orb}~
+~A_k^{nSH}\cos \phi_k^{nSH} ,
\eeq 

\noindent 
which involves six free parameters: the three amplitudes and three zero-points 
of {\it overtone} phases. 
In what follows we assume that those parameters remained constant during two nights. 

Since the light curves were based on time intervals $\Delta t=3P_{SH}=0.0365$d 
which are short compared to the two beat periods: 
$P_{orb/SH}=0.5571$d and $P_{nSH/SH}=0.3092$d we can replace $\phi_k^{orb}$ and 
$\phi_k^{nSH}$ in Eq.(4) with $\phi_k^{SH}$ and two, practically {\it constant} 
beat phases: $\phi_k^{orb}=\phi_k^{SH}+\phi_k^{orb/SH}$ and  
$\phi_k^{nSH}=\phi_k^{SH}+\phi_k^{nSH/SH}$. 

The resulting expression for $\Delta m_k$ depends now only on $\phi_k^{SH}$ 
and the phase of minimum $\phi_{k,min}^{SH}$ can be obtained from the condition 

\beq
d\Delta m_k/d \phi_k^{SH}=0 , 
\eeq

\noindent
while the observed amplitude of the light curve is  

\beq
A_{k,obs}~=~\Delta m_k(\phi_{k,min}^{SH}) . 
\eeq

Recalling that the phases of minimum $\phi_{k,min}$, determined  
in Section 3.2 from the Fourier analysis using Eq.(1), refer to the 
{\it fundamental} mode we must convert the {\it overtone} phases of minimum 
also to the {\it fundamental mode} phases by using Eq.(2). 

Finally, we determine the six unknown parameters, i.e. $A_k^{SH}$, $A_k^{orb}$, 
$A_k^{nSH}$, and $\phi_{k,min}^{SH}$, $\phi_{k,min}^{orb}$, $\phi_{k,min}^{nSH}$, 
by fitting -- via the least squares solution -- 
the amplitudes and phases of minimum, obtained from the procedure described above, 
to their observed values shown in the right panels of Figs 1 and 2. 
Results are listed in Table 1 under "1974", the formal errors being: 
$\sigma_A\sim \pm 0.6-1.5$mmag for $k=0$ and 1 and $\sigma_A\sim \pm 0.5-0.8$mmag 
for $k=2$ and 3, and $\sigma_{\phi}\sim \pm 0.02-0.04$. 
Shown in the right panels of Figs 1 and 2 are lines calculated with those parameters; 
they fit the points quite well (the only significant exception being $\phi_{2,min}$ 
in Fig.1). 
This implies that the parameters of the Fourier components did not change 
significantly during two nights. 

We now compare the amplitudes of the Fourier components with those obtained from 
the seasonal mean light curves (Section 2). 

(1) The amplitude of the superhump fundamental mode $A_0^{SH}=7.1$mmag was 
surprisingly large. This shows that this mode, commonly considered negligible, may 
occasionally contribute significantly to the observed light curve. 

(2) The amplitudes of the orbital overtones $A_1^{orb}=8.2$ and $A_2^{orb}=4.8$mmag 
were large, larger than that of the fundamental mode. This means that the shape 
of the orbital light curve may occasionally differ considerably from a simple 
cosine-wave. A good illustration is provided by the mean orbital light curve observed 
by Solheim et al. (1998, Fig.10) during their two-week WET campaign in 1990. 

(3) The amplitude of the negative superhump first overtone $A_1^{nSH}=4.4$mmag 
was comparable to that of the fundamental mode ($A_0^{nSH}=5.8$mmag) which suggests 
that the shape of the negative superhump light curve may occasionally differ 
considerably from a simple cosine-wave. Worth adding is that the amplitude 
of negative superhumps is highly variable (see Provencal et al. 1995, Fig.6).

\section { Conclusions } 

It has been rather obvious that large variations in the shape of the observed light 
curve of AM CVn are due to the interference between its three components: the superhumps, 
the negative superhumps and the orbital variations. 
Results presented above show that the shapes of the light curves of those three 
components are also variable and this contributes significantly to the observed 
variations. 

The time scale of those variations could be determined only from detailed analysis 
of long series of observations covering several consecutive nights; 
our results suggest only that it is longer that 1 day.

\begin {references} 

\refitem {Harvey, D.A., Skillman, D.R., Kemp, J., Patterson, J., Vanmunster, T., 
          Fried, R.E., Retter, A. } {1998} {\ApJ} {493} {L105} 

\refitem {Kotko, I., Lasota, J.-P., Dubus, G., Hameury, J.-M.} {2012} 
         {\AA} {544} {A13}

\refitem {Nelemans, G.} {2005} {\it ASP Conference Series} {330} {27}

\refitem {Nelemans, G., Steeghs, D., Groot, P.J.} {2001} {\MNRAS} {326} {621}

\refitem {Patterson, J., Halpern, J., Shambrook, A.} {1993} {\ApJ} {419} {803}

\refitem {Provencal, J.L. {\it et al.}} {1995} {\ApJ} {445} {927}

\refitem {Roelofs, G.H.A., Groot, P.J., Nelemans, G., Marsh, T.R., Steeghs, D.}
          {2006} {\MNRAS} {371} {1231}

\refitem {Skillman, D.R., Patterson, J., Kemp, J., Harvey, D.A., Fried, R.E., 
          Retter, A., Lipkin, Y., Vanmunster, T. } {1999} {\PASP} {111} {1281}

\refitem {Smak,J.} {1967} {\Acta} {17} {255}

\refitem {Smak,J.} {1975} {\Acta} {25} {372}

\refitem {Smak,J.} {2016} {\Acta} {66} {75} 

\refitem {Solheim, J.-E. et al.} {1998} {\AA} {332} {939} 

\end {references}

\end{document}